\documentclass[preprint, aps, prd, groupadress, nofootinbib, showkeys]{revtex4-1}

\usepackage{amsmath, amsfonts, amssymb, mathrsfs}
\usepackage{amsthm}
\usepackage{bbm}
\usepackage{cmbright}

\usepackage{graphicx}
\usepackage{float}
\usepackage[font=small]{caption}

\usepackage[hyperindex]{hyperref}
\hypersetup{breaklinks=true, hyperfootnotes=true, pagecolor=white, colorlinks=true, 
allcolors=blue
}
\usepackage[all]{hypcap}

\begin{document}

\title{Repulsive gravity induced by a conformally coupled scalar field implies a bouncing radiation-dominated universe}

\author{V. Antunes}\email[Electronic address:]{antunes@cbpf.br}
\author{M. Novello}\email[Electronic address:]{novello@cbpf.br}

 \affiliation{Centro Brasileiro de Pesquisas F\'{i}sicas (CBPF), Rua Dr. Xavier Sigaud 150, Urca, CEP 22290-180,\\
 Rio de Janeiro, RJ, Brazil}

\begin{abstract}

In the present work we revisit a model consisting of a scalar field with a quartic self-interaction potential non-minimally (conformally) coupled to gravity \cite{NOVELLO_1980}. When the scalar field vacuum is in a broken symmetry state, an effective gravitational constant emerges which, in certain regimes, can lead to gravitational repulsive effects when only ordinary radiation is coupled to gravity. In this case, a bouncing universe is shown to be the only cosmological solution admissible by the field equations when the scalar field is in such a broken symmetry state. 

\end{abstract}


\maketitle

\section{Introduction}

 Repulsive gravitational effects are
an essential ingredient of the Standard Cosmological Model (SCM), where they are evoked to overcome the difficulties faced by the Friedmann model, such as the horizon, flatness, and origin of the primordial density fluctuations problems \cite{GUTH_1981, LINDE_1982, ALBRECHT_STEINHARDT_1982, LINDE_1983, LINDE_2005}, as well as the observed late-time accelerated expansion of the universe \cite{RIESS_1998, PERLMUTTER_1999}. According to General Relativity (GR), when the minimal coupling principle is enforced, repulsive gravitational effects can only be generated by fluids with negative pressure. Such fluids are invariably behind inflationary models (see \cite{LINDE_2005} and references therein) and dynamical dark energy models \cite{DE_PDG_2013, COPELAND_2006, FRIEMAN_TURNER_HUTERER_2008}.

 A series of attempts to overcome the cosmological singularity problem, arguably the most severe difficulty faced by the Friedmann model (see \cite{NOVELLO_BERGLIAFFA_2008} and references therein), were proposed in the past in which the alternative concept of repulsive gravitational effects generated by fields non-minimally coupled to gravity was introduced \cite{LINDE_1979,NOVELLO_SALIM_1979, NOVELLO_1980}. In \cite{LINDE_1979}, in particular, a conformally coupled Higgs-like field with a potential $V(\phi) = m^2\phi^2 + \sigma\phi^4$ ($\sigma >0$) was responsible for a reversal of the effective gravitational constant in the early universe, when the density reached a critical value $\rho_c$. Such model was later ruled-out on stability grounds, since as $\rho \to \rho_c$ the effective gravitational constant diverges, $\kappa_{eff} \to \infty$ \cite{STAROBINSKY_1981}. Despite that, this idea paved the way for inflationary models based on non-minimally coupled Higgs-like scalar fields \cite{SALOPEK_BOND_BARDEEN_1989, FAKIR_UNRUH_1990, KAISER_1995, KOMATSU_FUTAMASE_1999, BEZRUKOV_SHAPOSHNIKOV_2008, BARVINSKY_KAMENSHCHIK_STAROBINSKY_2008} (see \cite{LINDE_2005} and references therein), and was the source of other gravitational repulsion models (see \cite{HOHMANN_WOHLFARTH_2010} and references therein). As we shall see, that criticism does not apply here.

 In the present work we revisit a model \cite{NOVELLO_1980}
in which ordinary matter (radiation) can generate repulsive gravitational effects by intervention of a conformally coupled scalar field. The main idea can be synthesized in the following steps:

\begin{itemize}
 \item[(i)] We assume the existence of a scalar field $\phi$ which has a quartic self interaction potential with the form $V(\phi) = m^2\phi^2 -\sigma \phi^4$ (we call attention to the relative sign);
 \item[(ii)] This scalar field couples non-minimally to gravity;
 \item[(iii)] We select among all possible candidates of such a coupling a conformal one;
 \item[(iv)] Matter interacts with the scalar field only through gravity;
 \item[(v)] The symmetry of the theory is broken by a state $\phi = \phi_0 = $ constant;
 \item[(vi)] When the field is in the broken symmetry state, the net result
on the equation for the metric is to produce an effective gravitational constant which, under certain conditions, has the opposite sign of the bare Newton's
gravitational constant, provided that the energy-momentum tensor for the matter fields is traceless ($T=0$).
 \item[(vii)] It will be shown that a necessary consequence of this process is that a ordinary radiation ($T=0$) dominated universe exhibits a bounce. 
\end{itemize}

\section{The model}

 \subsection{Non-minimally coupled scalar field}

In what follows, we adopt the space-time metric with signature $(+---)$.
The theory considered here is defined by a Lagrangian with a scalar field non-minimally (conformally) coupled to gravity 
\begin{equation}
L = \sqrt{-g}\left[ \frac{1}{\kappa}R + g^{\alpha\beta}\partial_{\alpha}\phi^{\ast}\partial_{\beta}\phi - V(\phi^{\ast}\phi)
- \frac{1}{6}R\phi^{\ast}\phi + \mathcal{L}_m \right],
\label{lagrangian}
\end{equation}
where $g_{\alpha\beta}$ are the components of the metric field, $g = \mbox{det}(g_{\alpha\beta})$, $R$ is the Ricci scalar,
$\kappa = 8\pi G$ is the reduced gravitational constant ($c=\hslash = 1$), $\mathcal{L}_m$ is the matter Lagrangian density,
and the self-interaction potential of the scalar field is given by \cite{NOVELLO_1980}
\begin{equation}
V(\phi^{\ast}\phi) = m^2\phi^{\ast}\phi - \sigma(\phi^{\ast}\phi)^2 - 2V_0.
\end{equation}
We call attention for the fact that this potential differs from the usual ones employed in \cite{LINDE_1979} or in inflation models (see \cite{LINDE_2005}) by the relative sign between the mass term, $m^2\phi^{\ast}\phi$, and the quartic self-interaction term, $\sigma(\phi^{\ast}\phi)^2$. 

The set of field equations obtained from the Lagrangian (\ref{lagrangian}) are
\begin{widetext}
\begin{subequations}
 \begin{align}
   R_{\alpha\beta} - \frac{1}{2}Rg_{\alpha\beta} = & -\frac{1}{{\frac{1}{\kappa} - \frac{1}{6}\phi^2}}\left[ \tau_{\alpha\beta}(\phi) + T_{\alpha\beta} \right],
\label{grav_field_eq}
 \end{align}
\begin{equation}
 \Box\phi + m^2\phi - 2\sigma\phi^3 + \frac{1}{6}R\phi = 0, \label{scalar_field_eq}
\end{equation}
\end{subequations}
\end{widetext}
where $\Box \equiv g^{\alpha\beta}\nabla_{\alpha}\nabla_{\beta}$, $\phi^2 \equiv \phi^{\ast}\phi$, $\phi^3 = (\phi^{\ast}\phi)\phi$, $T_{\alpha\beta}$ is the energy-momentum tensor of the matter fields, and
\begin{widetext}
\begin{align}
 \tau_{\alpha\beta}(\phi) \equiv \partial_{\alpha}\phi^{\ast}\partial_{\beta}\phi
- \frac{1}{2}\Big( \partial_{\mu}\phi^{\ast}\partial^{\mu}\phi - m^2\phi^2 + \sigma\phi^4 \Big)g_{\alpha\beta}
- \frac{1}{6}\Big( \nabla_{\alpha}\nabla_{\beta}\phi^2 - \Box\phi^2g_{\alpha\beta} \Big) - V_0 g_{\alpha\beta} \label{em_scalar}
\end{align}
\end{widetext}
is the ``improved" energy-momentum tensor of the scalar field.
Taking the trace of the field equation (\ref{grav_field_eq}),
and using equation (\ref{scalar_field_eq}), it follows that the Ricci scalar can be expressed as follows
\begin{equation}
 R = m^2\phi^2 - 4V_0 + T,
\end{equation}
where $T\equiv g^{\alpha\beta}T_{\alpha\beta}$. This enables us to rewrite equation (\ref{scalar_field_eq}) in the form
\begin{equation}
  \Box\phi + \left( m^2 - \frac{2}{3}V_0 + \frac{1}{6} T\right)\phi + \left(\frac{1}{6} m^2 - 2\sigma\right)\phi^3 = 0. \label{scalar_field_eq_b}
 \end{equation}

 \subsection{Broken symmetry and gravitational repulsion from ordinary matter}

It is clear from equation (\ref{grav_field_eq}) that in the case where the ground state of the scalar field potential vanishes, $V_0=0$,
the improved energy-momentum tensor of the scalar field $\tau_{\alpha\beta}(\phi)$
is irremediably not conserved. Following Callan et al. \cite{CALLAN_COLEMAN_JACKIW_1970}, however, we can define the conserved energy-momentum tensor
\begin{equation}
 E_{\alpha\beta}(\phi) \equiv \frac{1}{{\frac{1}{\kappa} - \frac{1}{6}\phi^2}}\tau_{\alpha\beta}(\phi). \label{em_complex}
\end{equation}
We now look for a constant solutions of equation (\ref{scalar_field_eq_b}) 
which correspond to stable vacua of the scalar field. Clearly, equation (\ref{scalar_field_eq_b}) only admits a constant solution in the special case where the trace of the energy momentum-tensor of matter fields is a constant. 
We stress the fact that the symmetry breaking process is only possible in this case, 
otherwise no nontrivial ground state solution of equation (\ref{scalar_field_eq_b}) exists.
Let us concentrate, from now on, on
matter fields described by a traceless energy-momentum tensor, $T=0$.

\begin{figure}[H]
   \centering
 \vspace{15mm}
  \includegraphics[scale=
0.8 
]{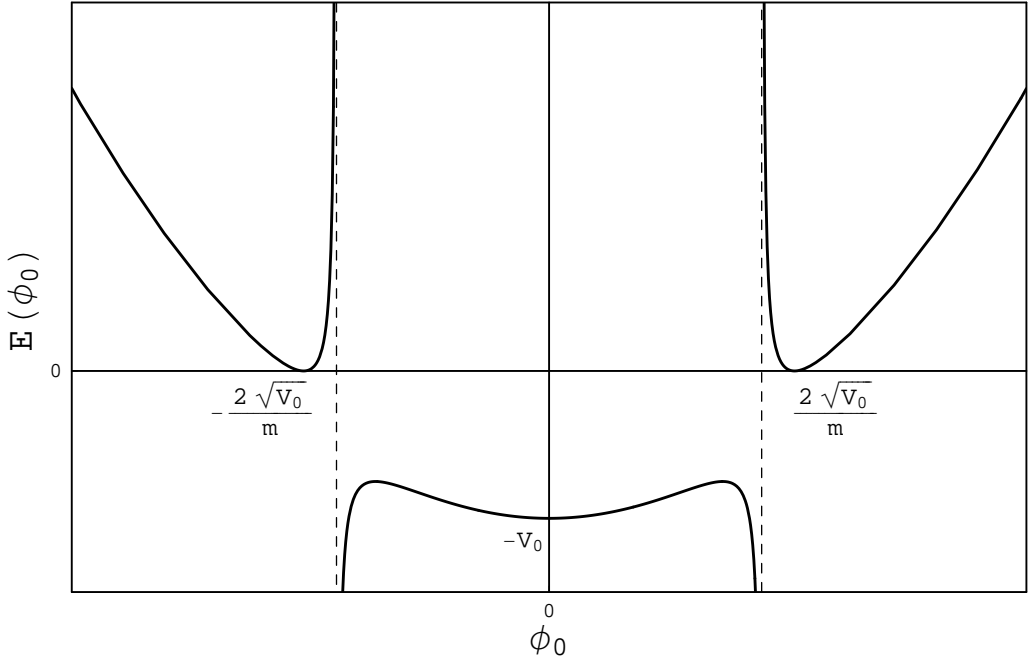} 
   \caption{Plot of the energy density $E(\phi_0)$, for $\kappa V_0>3m^2/2>0$.
The figure shows the solutions $\phi_0 = \pm 2\sqrt{V_0}/m$ corresponding to the nontrivial stable vacua,
and also two asymptotes at $\phi_0 = \pm \sqrt{6}$ separating three uncommunicating regionsa are indicated (dashed lines).} \label{energy_fig}
   \end{figure}

According to expressions (\ref{em_complex})-(\ref{em_scalar}), for a constant solution $\phi=\phi_0$ the energy density $E_{00}(\phi) \equiv E(\phi)$
corresponding to the energy-momentum tensor (\ref{em_complex}) has the form
\begin{equation}
E(\phi_0) = \frac{1}{{\frac{1}{\kappa} - \frac{1}{6}\phi_0^2}}(m^2\phi_0^2 - \sigma\phi_0^4 - 2V_0). \label{energy_func}
\end{equation}
On the other hand, for $T= 0$ equation (\ref{scalar_field_eq_b}) admits, besides the trivial solution $\phi_0 = 0$, two constant solutions
\begin{equation}
 \phi_0^2 = \frac{6m^2 - 4V_0}{12\sigma - m^2}. \label{relation}
\end{equation}
The constant nontrivial solutions which minimizes the energy density functional (\ref{energy_func}), and satisfies relation (\ref{relation}),
are given by
\begin{equation}
 \phi_0 = \pm \frac{2\sqrt{V_0}}{m}, \ \ \ \  \sigma = \frac{m^4}{8V_0}. \label{const_sol}
\end{equation}
The resulting behavior of the energy density $E(\phi_0)$ exhibits three uncommunicating regions, two of them containing the stable local minima
at $\phi_0 = \pm 2\sqrt{V_0}/m$, to which correspond $E(\phi_0) = 0$, the other
containing two unstable maxima and a metastable minimum at $\phi_0 = 0$ (see Fig \ref{energy_fig}).


Consequently, when the energy-momentum tensor for the matter fields is traceless ($T=0$), and the field is in a non-trivial stable ground state (\ref{const_sol}), the gravitational field equation (\ref{grav_field_eq}) assumes the form
\begin{align}
  R_{\alpha\beta} = -\kappa_{eff}T_{\alpha\beta} \label{grav_field_eq_2},
 \end{align}
where
\begin{equation}
\kappa_{eff} \equiv \frac{1}{\frac{1}{\kappa} - \frac{1}{6}\phi_0^2} =  \left(\frac{3m^2\kappa}{3m^2 - 2\kappa V_0}\right)
\end{equation}
is the renormalized gravitational constant.
The term multiplying the energy-momentum tensor of matter (ordinary radiation) in equation (\ref{grav_field_eq_2}) can, thus,
be viewed as a ``renormalized'' gravitational constant. Clearly, in the regime where the relation
\begin{equation}
\kappa V_0 > \frac{3}{2}m^2 > 0 \label{condition_mass}
\end{equation}
is satisfied ({\it i.e.} either the mass $m$ of the scalar field has to be very small, or $V_0$ has to be very large), gravity is reversed, {\it i.e.} $\kappa_{eff} < 0$. 

\section{Bouncing cosmological model}

  \subsection{Scalar field induced bounce}

We now investigate what cosmological solutions are compatible with the model discussed above. We assume spatial homogeneity and isotropy, and adopt a Friedmann metric for the space-time
\begin{equation}
ds^2 = dt^2 - a^2(t)\left[ \frac{dr^2}{1- \epsilon r^2} + r^2 \big(d\theta^2 + \sin^2\theta d\varphi^2\big) \right], \label{friedmann_metric}
\end{equation}
where $a(t)$ is the scale factor and $\epsilon$ determine the geometry of the spatial section. In order to be the source of the Friedmann metric (\ref{friedmann_metric}) the scalar field must depend on the cosmic time only, $\phi = \phi(t)$. We consider a radiation-dominated universe, $\rho = 3p$ ($T= 0$).
 In the regime $\phi = \phi_0$ the field equations reduce to
\begin{subequations}\label{cosm_eq}
\begin{equation}
\left( \frac{\dot{a}}{a} \right)^2 + \frac{\epsilon}{a^2} = \frac{1}{3}\kappa_{eff} \rho, 
\end{equation}
\begin{equation}
\frac{\ddot{a}}{a} = -\frac{1}{6}\kappa_{eff}(\rho + 3p), 
\end{equation}
\end{subequations}
where $\rho(t)$ and $p(t)$ are the density and pressure of the radiation fluid. In this case, if condition (\ref{condition_mass}) holds, gravity is reversed, $\kappa_{eff}<0$, and the field equations (\ref{cosm_eq}) only admit a solution for an open spatial section $\epsilon = -1$. Form energy conservation we have $\rho(t) \propto \rho_0 a^{-4}(t)$, where $\rho_0$ is a constant, so that the system of equations (\ref{cosm_eq}) yield
\begin{equation}
\dot{a} = \sqrt{ 1 - \frac{1}{3}|\kappa_{eff}| \rho_0 a^{-2}}.
\end{equation}
This equation can be readily integrated, and we obtain the following form for the scale factor
\begin{equation}
a(t) = \sqrt{ t^2  + a_0^2}. \label{scale_fac_bounce}
\end{equation}
Therefore, the universe in this model exhibits a bounce around $t = 0$, the constant $a_0 = \sqrt{|\kappa_{eff}| \rho_0/3}$ being the minimum value attainable by the scale factor. 

Interestingly, the bouncing solution obtained above coincides with a model based on a non-minimally coupled vector field proposed by one of the authors \cite{NOVELLO_SALIM_1979}, even though the two models differ considerably (see appendix A).

%
%

\section{Final Remarks}

In the present work we revisited a model \cite{NOVELLO_1980} in which a scalar field conformally coupled to gravity can generate
repulsive gravitational effects when only ordinary matter with traceless energy-momentum tensor (radiation) is coupled to gravity. It was shown that, in a radiation-dominated universe, when the scalar field is in a non-trivial ground (broken symmetry) state the only solution admissible by the field equations is a bouncing universe.

 \begin{appendix}
  
 \section{Bouncing model generated by a vector field non-minimally coupled to gravity \label{app_ns}}

 We include here a short review of the bouncing cosmological model proposed in \cite{NOVELLO_SALIM_1979} and the one presented above for an easy comparison between them. The Lagrangian describing a vector field non-minimally coupled to gravity employed in \cite{NOVELLO_SALIM_1979} is
 \begin{equation}
 L = \sqrt{-g}\left[ \frac{1}{\kappa}R + \beta RA_{\mu}A^{\mu} - \frac{1}{4}F_{\mu\nu}F^{\mu\nu} \right],
 \label{lagrangian_d}
 \end{equation}
where $\beta$ is a constant, and $F_{\mu\nu} = \nabla_{\mu}A_{\nu} - \nabla_{\nu}A_{\mu}$. The field equation are
\begin{widetext}
\begin{subequations} \label{field_eq_app}
 \begin{align}
   R_{\alpha\beta} - \frac{1}{2}Rg_{\alpha\beta} = & -\frac{1}{{\frac{1}{\kappa} + \beta A_{\mu}A^{\mu}}}\tau_{\alpha\beta}(A^{\mu}) ,
 \end{align}
\begin{equation}
 \nabla_{\nu}F^{\mu\nu} + \beta A^{\mu} = 0, 
\end{equation}
\end{subequations}
\end{widetext}
where $\tau_{\alpha\beta}(A^{\mu})$ is the improved energy-momentum tensor of the vector field. 

In a Friedmann geometry determined by the metric (\ref{friedmann_metric}), making the choice $A_{\mu} = A(t)\delta^0_{\mu}$, and defining $\Omega(t) = \frac{1}{\kappa} + \beta A^2(t)$, the set of field equations (\ref{field_eq_app}) assume the form
\begin{subequations}
\begin{equation}
3\frac{\ddot{a}}{a} = -\frac{\ddot{\Omega}}{\Omega},
\end{equation}
\begin{equation}
\frac{\ddot{a}}{a} + 2\left( \frac{\dot{a}}{a} \right)^2 + 2\frac{\epsilon}{a^2} = -\frac{\dot{a}}{a}\frac{\dot{\Omega}}{\Omega},
\end{equation}
\begin{equation}
\Box\Omega = 0.
\end{equation}
\end{subequations}
A particular solution for an open spatial section, $\epsilon = -1$, furnishes a scale factor with the form
\begin{equation}
a(t) = \sqrt{ t^2  + a_0^2},
\end{equation}
where $a_0$ is the minimum value of the scale factor. Although this is the same solution (\ref{scale_fac_bounce} ) obtained for the scalar field, the two models differs in crucial points. First, the quantity $\Omega(t)$, which is analogous to the effective gravitational constant of the model discussed in the main text, is not a constant but a function of the cosmic time (although a model similar to the one considered in the main text would arise in the special case $A_{\mu}A^{\mu} =$ constant). Second, here the non-minimally coupled vector field alone can be the source of the geometry, while in the bouncing model induced by the scalar field, in the absence of matter fields and for the scalar field in the non-trivial (broken symmetry) ground state, the gravitational field equations reduce to the vacuum Einstein equations.

\end{appendix}

\vspace{5mm}

\section*{Acknowlegments}

The authors would like to thank the Brazilian National Council of Technological and Scientific Development (CNPq) and the Research Support Foundation of the State of Rio de Janeiro (FAPERJ) for a grant.


\end{document}